\documentclass[preprint,3p]{elsarticle}
\makeatletter
\def\ps@pprintTitle{%
 \let\@oddhead\@empty
 \let\@evenhead\@empty
 \def\@oddfoot{}%
 \let\@evenfoot\@oddfoot}

\usepackage{amssymb}
\usepackage{amsmath}

\usepackage{verbatim}
\usepackage{threeparttable}
\usepackage{multirow}
\usepackage{lineno}
\usepackage{hyperref} 

\usepackage[colorinlistoftodos]{todonotes}

\newcounter{bla}

\journal{Computer Physics Communications}

\begin{document}

\begin{frontmatter}



\title{CRANE: A MOOSE-based Open Source Tool for Plasma Chemistry Applications}


\author[a]{Shane Keniley}
\author[a]{Davide Curreli}

\cortext[author] {\textit{E-mail addresses:} keniley1@illinois.edu (S. Keniley), dcurreli@illinois.edu (D. Curreli)}
\address[a]{Department of Nuclear, Plasma, and Radiological Engineering, University of Illinois at Urbana-Champaign, Urbana, IL, 61801}

\begin{abstract}
Numerical simulation of plasma discharges is often performed by models developed in-house and coupling externally and separately written codes. The MOOSE (Multiphysics Object Oriented Simulation Environment) framework provides tools for quickly developing and coupling together software in a scalable framework, making it well-suited for plasma simulations. A new MOOSE application, CRANE (Chemical ReAction NEtwork), was developed in order to add an independent chemical kinetics application to the framework. In this work the capabilities of CRANE are detailed, including its use as a standalone solver for global plasma chemistry models. The capability of fully coupling applications in the MOOSE framework is also shown by compiling CRANE directly into the MOOSE-based plasma transport software Zapdos, showing the possibility to solve fully coupled drift-diffusion-reaction plasma chemistry problems. The code is open-source and available at the following url, \url{https://github.com/lcpp-org/crane}.  
%
%
%

\end{abstract}


\end{frontmatter}

\section{Introduction}
\label{sec:introduction}

Interest in low temperature plasma science, and in particular plasma chemistry, has begun to increase further in recent years due to its possible applications in a wide variety of fields such as medicine \cite{Ratovitski2014}, agriculture \cite{Ito2018}, and chemical production \cite{Zhang2017a}. Plasma simulations often require a nonlinear and nonequilibrium system of equations to be solved, and including a sufficient number of chemical reactions is vital for accurately modeling a discharge. Furthermore, the problem becomes a multiscale and multiphysics system that requires the coupling of different regimes when plasma-material interactions are necessary, which often means either stitching together multiple codes developed with varying languages and standards, using models of limited portability \cite{Xiong2012}, or purchasing a COMSOL license.

In this work we introduce CRANE, an open-source plasma chemistry software developed within the MOOSE framework that was designed to address nonequilibrium plasma chemistry problems of arbitrary size. The Multiphysics Object Oriented Simulation Environment (MOOSE) is a finite element software developed at Idaho National Laboratory \cite{Gaston2009} that was created to provide a modular software framework for  coupled nonlinear multiphysics simulations. Its modular structure allows different codes to be coupled in multiple different ways: they may be compiled together and included in a single fully coupled nonlinear solver, or applications can be loosely coupled with the \texttt{MultiApp} system \cite{Gaston2015}. The environment has previously been used to study a wide variety of physical applications such as superconductivity \cite{Mangeri2015}, incompressible Navier-Stokes systems \cite{Peterson2018}, and large-scale nuclear reactor simulations \cite{Gaston2015}. The framework is well suited to the study of plasmas due to its massively parallel operation and its ability to natively treat multi-phase systems (such as plasma-liquid interactions). Indeed, development of a suite of plasma-relevant software has begun on the platform with the plasma transport model, Zapdos, which was created and used to study low temperature plasma discharges on liquid water \cite{Lindsay2016}. 

CRANE aims to further expand the plasma simulation capabilities of the MOOSE framework by including a chemical kinetics solver. Using the MOOSE framework's \textit{Actions} system, CRANE allows lists of chemical reactions to be added to the input file in a simple human-readable format to construct reaction networks. It may be run alone as a 0D model or it can leverage the coupling capabilities of the framework and be compiled into other MOOSE applications as a submodule, providing separately-developed software with the capability to solve reaction networks. Using CRANE alone requires no C++ programming from the user, and directly coupling applications requires only modifying the main application's Makefile and adding two lines to the applications main C++ file. In this way all of the functionality of CRANE becomes available to the parent application through the text-based input files used by MOOSE applications. 

The ability to couple different codes together represents a significant advantage both for CRANE and for MOOSE applications in general. By compiling Zapdos and CRANE together, a user is able to simulate a fully-coupled system of multispecies drift-diffusion-reaction equations. A large number of reactions may be easily added to a Zapdos-CRANE simulation without writing any code, allowing a user to model a plasma discharge with necessarily large reaction networks, such as nitrogen and oxygen plasmas \cite{Flitti2009}. Since the dimensionality and parallel capabilities of a problem are handled internally by MOOSE, a coupled Zapdos-CRANE model may be easily scaled into multiple dimensions and parallelized from the input file and command line. Future plasma-relevant software that is developed in the MOOSE framework will have the ability to be coupled together with Zapdos and CRANE as well.


\section{Software Description}
\label{sec:software}

\subsection{Chemical Kinetics}
\label{sec:kinetics}

CRANE was developed to study the problem of chemical kinetics in nonequilibrium, multi-fluid plasmas. The mathematical kernel of the software is a coupled system of ODE rate equations: 

\begin{equation}
\frac{d n_i}{dt} = \sum_{j=1}^{j_{max}} S_{ij}
\end{equation}

where $n_i$ is the concentration of species $i$, and the right hand side is the sum over all $j$ source and sink terms of that species. For a two-body reaction $j$ between species A, B, and C, with stoichiometric coefficients a$_1$, a$_2$, b, and c, and rate coefficient $k$:

\begin{equation}
a_1 A + bB \xrightarrow{k} a_2 A + cC
\end{equation}
The reaction has an associated reaction rate $R_j$: 

\begin{equation}
    R_j = k_j [A]^{a_1} [B]^b
\end{equation}
and the rate of production for each species may then be calculated from the reaction rate and stoichiometric coefficients: 

\begin{align}
S_A &= (a_2 - a_1)R_j \\
S_B &= -b k_j R_j \\
S_C &= c R_j
\end{align}
Under the adiabatic isometric approximation, the gas temperature may be changed as a result of each reaction's associated change in enthalpy, $\delta \epsilon_j$:

\begin{equation}
\frac{N_{gas}}{\gamma-1} \frac{dT}{dt} = \sum_{j=1}^{j_{max}} \pm \delta \epsilon_j R_j
\end{equation}
where $N_{gas}$ is the neutral gas density and $\gamma$ is the specific gas heat ratio. 

\subsection{User Interface}
\label{sec:moose}

An input file for CRANE largely follows the same syntax as all MOOSE applications. The input file requires at minimum five ``blocks" (Mesh, Variables, Kernels, Executioner, Outputs). Variables declares each nonlinear variable in the system (in CRANE's case, these are the species densities), Kernels represent a single term or piece of physics in the system of equations being solved, the Executioner dictates the numerical scheme and timestepping parameters, and the output file format is dictated in the Outputs block. When used alone, CRANE solves a global system of ODEs and uses only scalar variables (denoted by the \texttt{family = SCALAR} parameter) and ScalarKernels. 
The smallest possible mesh, a point, is used to ensure full compatibility with MOOSE.
An example input file is shown below. 

\begin{small}
\begin{verbatim}
[Mesh]
  type = GeneratedMesh
  dim = 1
  nx = 1
[]

[Variables]
  [./e]
    family = SCALAR
    order = FIRST
    initial_condition = 1
  [../]

  [./Ar]
    family = SCALAR
    order = FIRST
    initial_condition = 2.5e19
    scaling = 1e-19
  [../]

  [./Ar+]
    family = SCALAR
    order = FIRST
    initial_condition = 1
  [../]
[]

[ScalarKernels]
  [./de_dt]
    type = ODETimeDerivative
    variable = e
  [../]

  [./dAr_dt]
    type = ODETimeDerivative
    variable = Ar
  [../]

  [./dAr+_dt]
    type = ODETimeDerivative
    variable = Ar+
  [../]
[]

[ChemicalReactions]
  [./ScalarNetwork]
    species = 'e Ar Ar+'
    file_location = 'example_folder'
    sampling_variable = 'reduced_field'

    reactions = 'e + Ar -> e + e + Ar+   : EEDF (rxn1.txt)
                 e + Ar+ + Ar -> Ar + Ar : 1e-25'

   [../]
[]

[AuxVariables]
  [./reduced_field]
    order = FIRST
    family = SCALAR
    initial_condition = 50e-21
  [../]
[]


[Executioner]
  type = Transient
  end_time = 0.28e-6
  dt = 1e-9
  solve_type = NEWTON
  line_search = basic
[]

[Preconditioning]
  [./smp]
    type = SMP
    full = true
  [../]
[]

[Outputs]
  csv = true
  interval = 10
[]

\end{verbatim}
\end{small}

While in principle every term in the system of equations must be included as a Kernel (or ScalarKernel), CRANE was developed utilizing the \textit{Actions} system in the MOOSE framework to automatically add a system of reactions, which is shown in the \texttt{ChemicalReactions} block. This example adds six source and sink ScalarKernels to the solver automatically without requiring the user to individually add each term to the ScalarKernels block. In this example, the nonlinear species are named in the \texttt{species} parameter (electrons, neutral argon, and ionized argon), and the reactions (separated by a return character) are listed in the \texttt{reactions} parameter. Rate coefficients are separated from each reaction by a colon character. The first reaction's rate coefficient is indicated to be tabulated in a file named `rxn1.txt' located in the `example\_folder' directory. The \texttt{sampling\_variable} parameter dictates what such rate coefficients are tabulated with; in this case it is tabulated as a function of the \texttt{reduced\_field} parameter, which is an \texttt{AuxVariable} with a constant value of 50 Td.

As part of the MOOSE framework, CRANE has access to a wide array of options for tuning a simulation. Solver options such as numerical schemes, adaptive timestepping, and PETSc options are denoted in the \texttt{Executioner} block. MOOSE includes multiple explicit and implicit time integrators: available implicit methods are backward Euler, Crank Nicolson, BDF2, DIRK, and Newmark-$\beta$, while the available explicit methods are forward Euler, Midpoint, and total variation-diminishing Runge-Kutta second order method. Note that this is only intended to be a brief summary of options relevant to CRANE. A detailed list of all input file options are available on the MOOSE framework website: \url{https://mooseframework.inl.gov}

\subsection{Code Coupling}
\label{sec:coupling}

The largest advantage that CRANE has over similar chemistry solvers such as ZDPlasKin \cite{Pancheshnyi2008} and CHEMKIN \cite{Kee1996} is that it may be natively coupled to other separately-developed MOOSE applications, without requiring additional coding from the user. For example, when coupled to the low temperature plasma transport code, Zapdos, all of the functionality built into CRANE becomes natively accessible by Zapdos through the application's input file. No data transfer is necessary in this case since the codes are compiled together and treated as a single application. In this way the problem becomes a fully coupled system of drift-diffusion-reaction equations.

\section{Application Examples}
\label{sec:examples}

\subsection{Global Argon Chemistry Model}
\label{sec:global}

As a standalone solver, CRANE may be used to study global reaction networks in plasma discharges. In this example we model a microcathode discharge in argon \cite{Stark1999}. The model includes the same parameters as used in the ZDPlasKin example based on Pancheshnyi's model \cite{Pancheshnyi2008}: a pressure of 100 mTorr was set, the discharge gap was set to 4 mm, and a circuit including a 1 kV power supply and a 100 $k\Omega$ resistor were used to calculate rate coefficients and the reduced electric field. The reduced electric field was set as an \texttt{AuxVariable} and computed at the end of each timestep under a local field approximation based on the gas density N, applied voltage $V_{dc}$, gap distance $d$, and resistance $R$: 
\begin{equation}
\frac{E}{N} = \frac{V_{dc}}{dN + eRAn_e (\mu N)}
\end{equation}
where $A$ is the area of the electrode with radius $r = 4$ mm, $\mu$ is the electron mobility, $n_e$ is the electron density, and $e$ is the elementary charge.

\begin{figure}[h]
	\centering
	\includegraphics[width=0.6\textwidth]{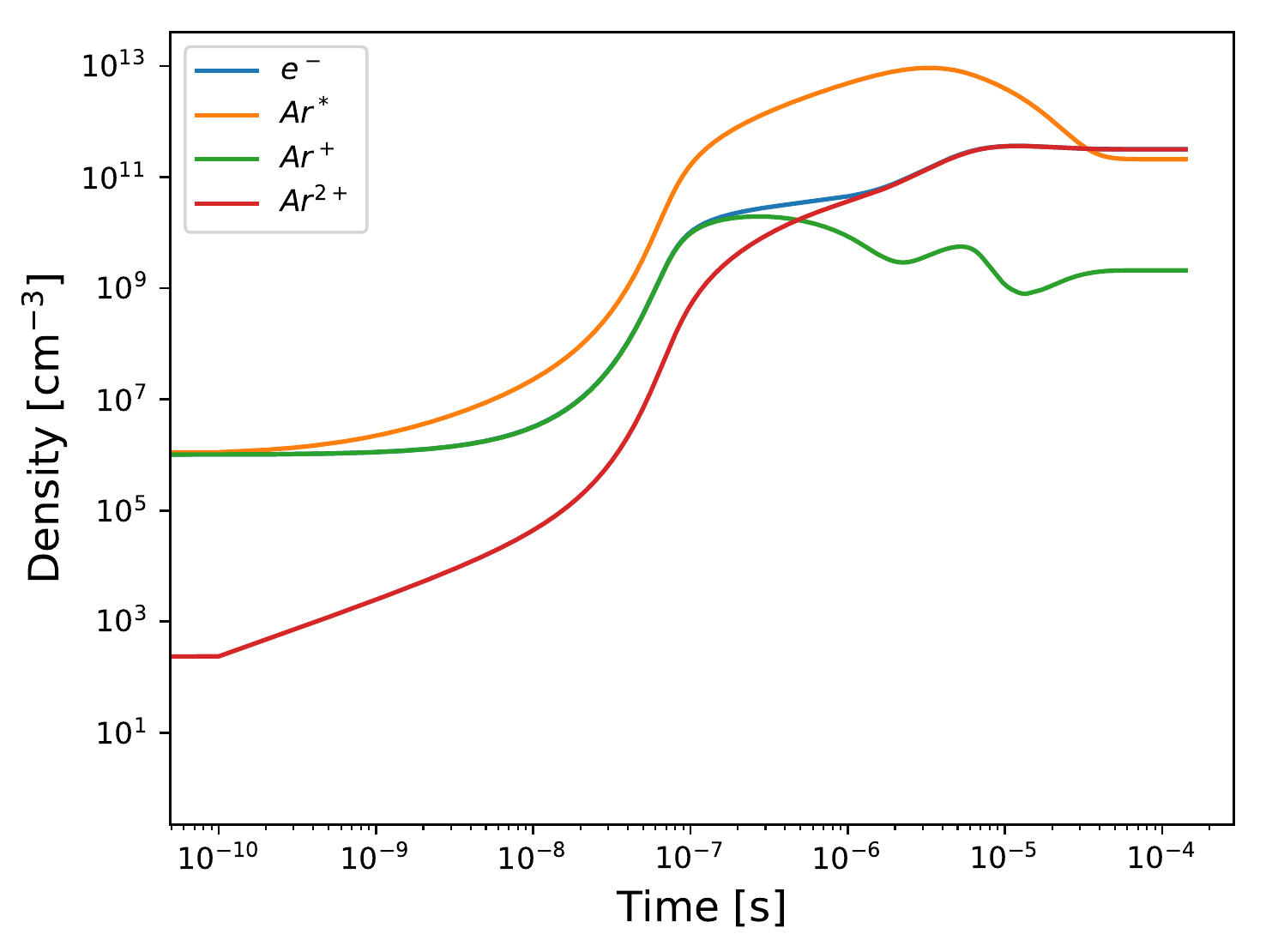}
	\caption{Density of three different argon species and electrons as a function of time. Test adapted from \cite{Pancheshnyia}}
	\label{fig:global}
\end{figure}

\begin{table}[h]
\begin{center}
 	\begin{threeparttable}
	\centering
	\caption{Reactions and rate coefficients included in global argon model. \cite{Pancheshnyi2008} }
	\renewcommand{\arraystretch}{1.2}
	\begin{tabular}{ l  c  r }
		\hline
 		Reaction & Rate Coefficient & Units   \\
 		\hline
 		$e + Ar \rightarrow 2e + Ar^+$ \quad & EEDF & cm$^{-3}$ s$^{-1}$ \\
 		$e + Ar \rightarrow Ar^* + e$ \quad & EEDF &   \\
 		$e + Ar^* \rightarrow Ar + e$ \quad & EEDF &   \\
 		$e + Ar^* \rightarrow Ar^+ + 2e$ \quad & EEDF &   \\
		$Ar^{2+} + e \rightarrow Ar^* + Ar$ \quad & $8.5 \times 10^{-7} (\tfrac{Te}{300})^{-0.67}$ &  \\
 		$Ar^{2+} + Ar \rightarrow Ar^+ + 2Ar$ \qquad & $\tfrac{6.06 \times 10^{-6}}{T_{gas}} \exp(-15130/T_{gas})$ &  \\
 		$Ar^* + Ar^* \rightarrow Ar^{2+} + e$ \quad & $6.0\times 10^{-10}$ &  \\ \hline
 		$Ar^+ + 2e \rightarrow Ar + e$ \quad & $8.75\times 10^{-27} (\tfrac{Te}{11600})^{-4.5}$ & cm$^{-6}$ s$^{-1}$ \\
 		$Ar^* + 2Ar \rightarrow 3Ar$ \quad & $1.4\times10^{-32}$ &  \\
 		$Ar^+ + 2Ar \rightarrow Ar^{2+} + Ar$ & $2.25\times 10^{-31} (\tfrac{T_{gas}}{300})^{-0.4}$ &  \\ \hline
 		$e \rightarrow e(W)$ \quad & $k_{diff}$* & s$^{-1}$ \\
 		$Ar^+ \rightarrow Ar^+(W)$ \quad & $k_{diff}$* & \\
 		$Ar^{2+} \rightarrow Ar^{2+}(W)$ \quad & $k_{diff}$* & \\
		\hline
	\end{tabular}
	\label{tab:global_argon}
	\begin{tablenotes}\footnotesize
	\item[*] $k_{diff} = 1.52 (\tfrac{760  \text{ Torr}}{P_{gas}})(\tfrac{T_{gas}}{273.16 K}) (\tfrac{T_e}{11600})( (\tfrac{2.405}{r})^2 + (\tfrac{\pi}{L})^2 )$
	\item[*] r = 4 mm (radius), L = 4 mm (gap length)
	\end{tablenotes}
	\end{threeparttable}
	\end{center}
\end{table}

Electron mobility, electron temperature, and the electron-impact rate coefficients must be computed from cross section data through convolution with an electron energy distribution function (EEDF). In the present work this is accomplished with the Bolsig+ software \cite{Hagelaar2005}. CRANE includes two separate methods of reading rate coefficients from external software: (1) rate coefficients may be tabulated prior to running crane as a function of either reduced electric field or electron temperature, or (2) Bolsig+ may be run by CRANE periodically throughout the simulation to recompute transport and rate coefficients, and the results will be automatically tabulated and read by CRANE. For this example Bolsig+ is run every 100 timesteps until the argon ion density begins to rapidly change at $t = 2 \mu s$, at which point the electron-impact rate coefficient tables are held constant for the remainder of the simulation. The full list of reactions and rate coefficients for the electrons and four argon species (Ar, Ar$^+$, Ar$^{2+}$, and Ar$^*$) are shown in Table~\ref{tab:global_argon}. There are also three surface loss ``species" to account for diffusion losses: e(W), Ar$^+$(W), and Ar$^{2+}$(W).

Fig.~\ref{fig:global} presents the time evolution of electrons, excited argon (Ar$^*$), and both argon ion densities (Ar$^+$, Ar$^{2+}$). All species densities begin to increase as the electrons collide with the neutrals to form excited and ionized species. After $t = 0.1$ ms the system achieves steady state. Note that as with the ZDPlasKin example, electrons and Ar$^+$ are started with a non-zero value ($10^6 cm^{-3}$) in order to initiate the discharge. All results show good agreement with both the ZDPlasKin example and the COMSOL example based on the same model.

\subsection{Collisional Radiative Model}
\label{sec:collisional_radiative}

\begin{figure}[h]
	\centering
	\includegraphics[width=0.6\textwidth]{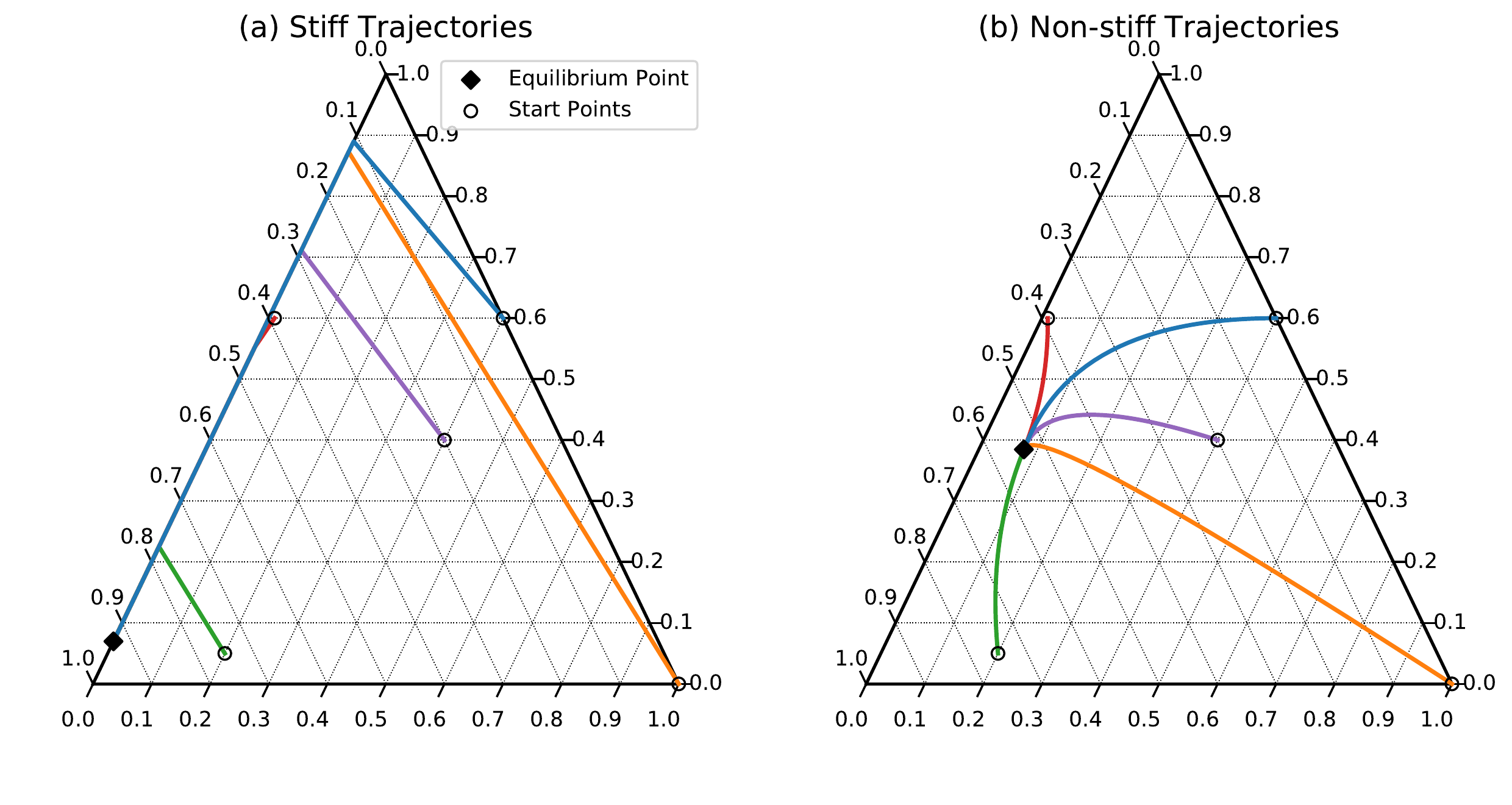}
	\caption{CRANE solution of a Collisional Radiative Model of a simple 3-level hydrogen atom \cite{Rehman2018}, for (a) stiff transition strengths and (b) non-stiff transition strengths. The axes of the ternary plots report the concentrations of each of the three states of the atom. Each plot is for five different initial concentrations; note that the initial conditions are the same between the two tests, and equal to (0.4, 0.6, 0.0), (1.0, 0.0, 0.0), (0.2, 0.05, 0.75), (0.01, 0.6, 0.39), (0.4, 0.4, 0.2).} 
	\label{fig:trajectories}
\end{figure}

Here we consider a simplified collisional radiative model of a three-level hydrogen system, adapted from Rehman's work \cite{Rehman2018}. Six transitions were considered in Rehman's model with three different species, corresponding to three coupled differential equations: 

\begin{align}
\frac{d n_1}{dt} &= -(f_{12} + f_{13})n_1 + f_{21}n_2 + f_{31} n_3 \\
\frac{d n_2}{dt} &= f_{12}n_1 - (f_{23} + f_{21})n_2 + f_{32}n_3 \\
\frac{d n_3}{dt} &= f_{13}n_1 + f_{23} n_2 - (f_{31} + f_{32})n_3
\end{align}

The transition frequences are shown in Table~\ref{tab:frequencies}. Indices 1, 2, and 3 refer to $H$, $H^*$, and $H^+$, respectively, corresponding to a system of three reversible reactions: 

\begin{align}
H + e &\leftrightharpoons H^* + e \label{eq:transition1} \\
H^* + e &\leftrightharpoons H^+ + 2e \label{eq:transition2} \\
H + e &\leftrightharpoons H^+ + 2e \label{eq:transition3}
\end{align}

In this example the system of reactions was solved with five different initial conditions, and the simulations were run in both the stiff and non-stiff cases. A CRANE input file will read equations written in the form of Eqs.~\ref{eq:transition1}-\ref{eq:transition3} in the \texttt{ChemicalReactions} block and automatically include the source and sink terms.  The input file is included in the CRANE GitHub repository as \texttt{example2.i}.  The implicit Newton-Krylov solver with an initial timestep of $dt = 10^{-8} s$ was used in both cases. Results of the simulation are shown in Fig.~\ref{fig:trajectories}. In the stiff case all of the trajectories first merge before reaching equilibrium, while in the non-stiff case each trajectory follows a unique path before converging at equilibrium. Both the trajectory behaviors and equilibrium points are the same as computed by Rehman.  

\begin{table}
	\centering
	\caption{Transition frequencies for both the stiff and non-stiff problems. \cite{Rehman2018}.}
	\begin{tabular}{ c  c  c }
		\hline
 		Frequency & Stiff [$s^{-1}$] & Non-stiff [$s^{-1}$] \\
 		\hline
 		$f_{12}$ & $2.7 \times 10^{10}$ & $9.0 \times 10^{1}$ \\
 		$f_{13}$ & $9.0 \times 10^{8}$ & $1.0 \times 10^{2}$\\
 		$f_{23}$ & $1.0 \times 10^{6}$ & $5.0 \times 10^{1}$\\
 		$f_{32}$ & $7.5 \times 10^{4}$ & $3.0 \times 10^{1}$\\
		$f_{21}$ & $3.8 \times 10^{1}$ & $1.0 \times 10^{1}$\\
 		$f_{31}$ & $1.7 \times 10^{2}$ & $2.0 \times 10^{1}$\\
		\hline
	\end{tabular}
	\label{tab:frequencies}
\end{table}

\subsection{Nitrogen Chemistry}
\label{sec:nitrogen}

\begin{figure}[h]
    \centering
    \includegraphics[width=0.5\textwidth]{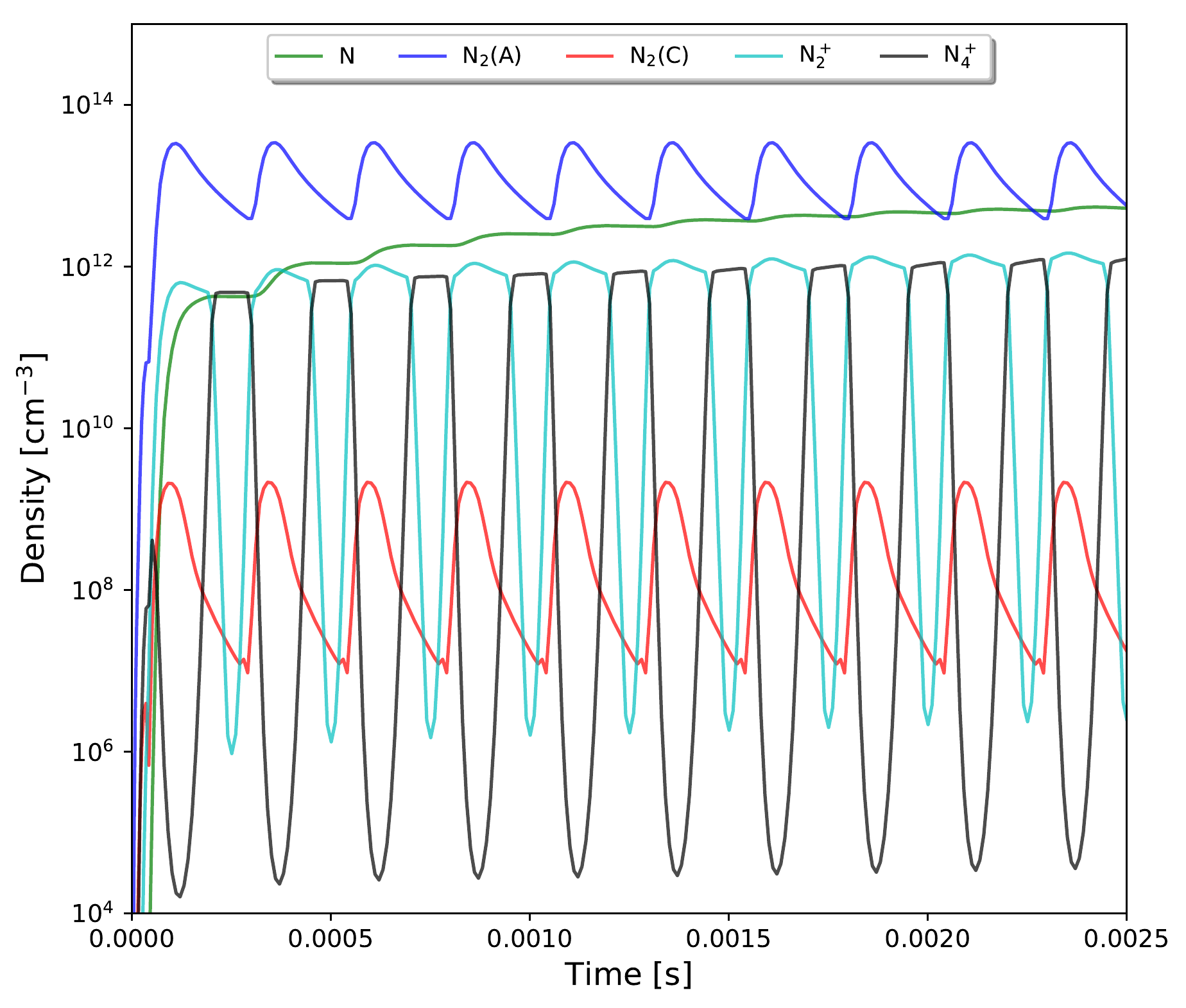}
    \caption{Five nitrogen species densities plotted as a function of time. Test adapted from \cite{Pancheshnyi}}
    \label{fig:nitrogen}
\end{figure}

This example utilizes experimentally measured electron density and electric field values as input to CRANE. Similarly to the model present in Sec.~\ref{sec:global}, this model follows the same species and procedure as a ZDPlasKin example \cite{Pancheshnyi}. The model includes 10 species and 36 reactions, with electron density and reduced electric field being read in from tabulated data as \texttt{AuxVariables}. This model is included in the CRANE GitHub repository as \texttt{example4.i}. The results of the simulation are shown in Fig.~\ref{fig:nitrogen}.
\begin{figure}[h!]
	\centering
	\includegraphics[width=0.8\textwidth]{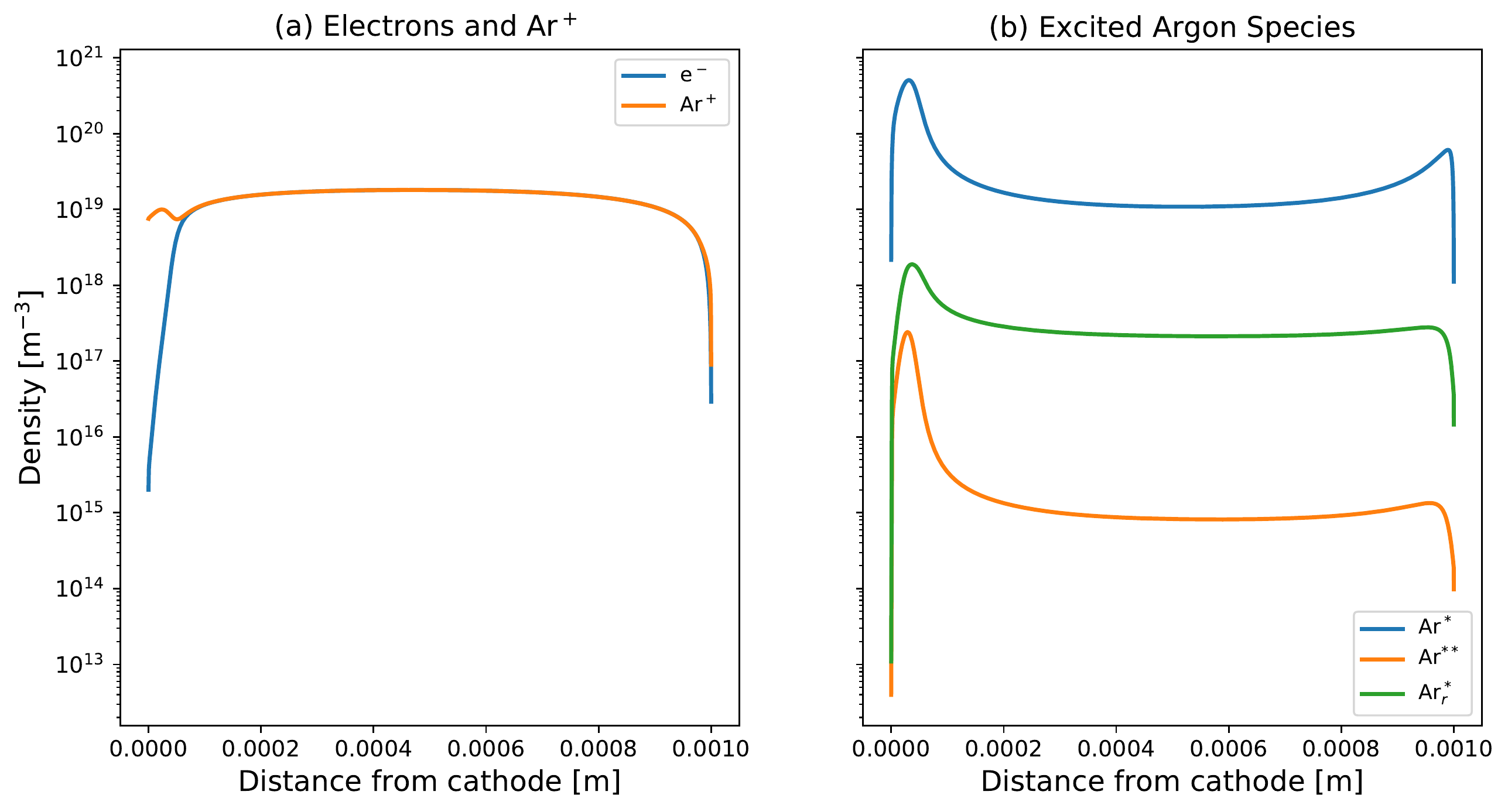}
	\caption{Density of electrons and argon ions (left) and excited argon species (right) across the domain at steady state. Cathode is on the left (L = 0), while a grounded wall is at L = 1mm.}
	\label{fig:zapdos_densities}
\end{figure}
\subsection{Fully-Coupled Plasma Chemistry}
\label{sec:zapdos}
In this section we apply CRANE's chemical reaction network to a 1D system by coupling to the plasma transport software Zapdos.  While MOOSE has a robust framework for coupling multiscale simulations through the \textit{MultiApp} system \cite{Gaston2015}, in this case the problem is not multiphysics; rather, it is a set of drift-diffusion-reaction equations that are fully coupled, and as such it must be computed on a single fully-coupled nonlinear solver. This is done by compiling CRANE directly into Zapdos as a submodule.

In this example, the model is a DC discharge applied to a pure argon environment at atmospheric pressure and room temperature, with a gap of $L = 1$ mm. The transport model in Zapdos is based on the drift-diffusion approximation, and Poisson's equation including all charged species is solved to calculate the electric field. The boundary conditions are the same as presented in Lindsay's paper \cite{Lindsay2016}, but with a grounded electrode at $L = 1mm$ rather than a liquid interface. Electrons and four argon species are included in this model, including ions (Ar$^+$) and three excited states (Ar$^*$, Ar$^{**}$, Ar$_r^*$), with a system of thirteen reactions being included. The list of reactions and rate coefficients is shown in Table~\ref{tab:zapdos_reactions}. Note that this Zapdos example uses molar densities, and as such the rate coefficients are presented in units of mol$^{-1}$ m$^{-3}$ s$^{-1}$. 

\begin{table}[h]
	\centering
	\caption{Reactions and rate coefficients included in global argon model. \cite{Eliseev2016} }
	\renewcommand{\arraystretch}{1.2}
	\begin{tabular}{ l  c  r }
		\hline
 		Reaction & Rate Coefficient & Units   \\
 		\hline
 		$e + Ar \rightarrow 2e + Ar^+$ \quad & EEDF & mol$^{-1}$ m$^{-3}$ s$^{-1}$ \\
 		$e + Ar \rightarrow Ar^* + e$ \quad & EEDF &   \\
 		$e + Ar \rightarrow Ar_r^* + e$ \quad & EEDF &   \\
 		$e + Ar \rightarrow Ar^{**} + 2e$ \quad & EEDF &   \\
		$e + Ar^* \rightarrow e + Ar_r^*$ \quad & $1.2044 \times 10^{11}$ &  \\
 		$Ar + Ar^{**} \rightarrow Ar + Ar^*$ \qquad & $2.4088 \times 10^7$ &  \\
 		$Ar^* + Ar^* \rightarrow Ar^+ + e$ \quad & $7.2264\times 10^8$ &  \\ 
 		$Ar^{**} + Ar^{**} \rightarrow Ar^+ + e$ \quad & $7.2264\times 10^8$ &  \\
 		$Ar_r^* + Ar_r^* \rightarrow Ar^+ + e$ \quad & $7.2264\times 10^8$ &  \\
 		$Ar^* + Ar^{**} \rightarrow Ar^+ + e$ & $7.2264\times 10^8$ &  \\ 
 		$Ar^* + Ar_r^* \rightarrow Ar^+ + e$ \quad & $7.2264\times 10^8$ &  \\ \hline
 		$Ar^{**} \rightarrow Ar^*$ \quad & $3.00 \times 10^7$ & s$^{-1}$ \\
 		$Ar_r^* \rightarrow Ar$ \quad & $3.33 \times 10^8$ & \\
		\hline
	\end{tabular}
	\label{tab:zapdos_reactions}
\end{table}

The excited level components, reactions, and rate coefficients are all adapted from Eliseev's work \cite{Eliseev2016}. Electron-impact rate coefficients and electron mobility and diffusivity were pre-tabulated as a function of electron temperature by Bolsig+, and the cross sections were found from the LXCat database \cite{BSRDatabase}. Steady state was achieved after 53 $\mu s$.  The simulation took 12 seconds to run with a single core on a 2015 MacBook Pro with a 2.7 GHz Intel Core i5 processor. Densities of the argon ions and electrons are shown in Fig.~\ref{fig:zapdos_densities}a, and excited argon species densities are shown in Fig.~\ref{fig:zapdos_densities}b. Electrons are depleted near the cathode and anode, as expected. The excited species all show a strong peak near the cathode.

\section{Code Benchmarking}
\label{sec:benchmarking}

The computational time required to solve a reaction network in CRANE is primarily dependent on the number of species and reactions. Four different reaction network tests with a varying number of species and reactions were run to model this dependence. The results were compared to the same reaction networks modeled by ZDPlasKin. 
All tests were run on a 2015 MacBook Pro with a 2.7 GHz Intel Core i5 processor with 8GB DDR3 memory. Since MOOSE simulations require a mesh to run, CRANE simulations were initialized with a 1D, single element ``dummy" mesh (\texttt{dim = 1}, \texttt{nx = 1}). MOOSE applications include many options to optimize a solver, including multiple preconditioners, PETSc options, and numerical schemes, all of which are highly problem-dependent and as such are not explored in this work. All CRANE simulations presented here were performed with the implicit Euler numerical scheme and a Newton-Krylov solver with a relative tolerance of $10^{-4}$, and no additional PETSc options were enabled. The variable \texttt{scaling} parameter was used to scale variable residuals to be closer to unity to decrease the number of nonlinear iterations necessary for convergence. Simulation timing is shown in Table~\ref{tab:crane_benchmark}.

\begin{table}[h]
	\centering
	\begin{threeparttable}
	\caption{Benchmarking CRANE against ZDPlasKin. Times are an average out of 10 runs.}
	\renewcommand{\arraystretch}{1.2}
	\begin{tabular}{ | c | c | c | c | c | c |}
		\hline
 		\multirow{2}{0.8cm}{\textbf{Test}} & \multirow{2}{2cm}{\textbf{Species}} & \multirow{2}{2cm}{\textbf{Reactions}} & 
 		 \multicolumn{2}{c|}{\textbf{Simulation Time} (s)} &
 		 \multirow{2}{1cm}{\textbf{Refs.}} \\
		\cline{4-5}  		
 		& & & \textit{CRANE} & \textit{ZDPlasKin} &  \\ \hline
 		1 & 2 & 2 & 0.410 & 0.0192 & \cite{Pancheshnyib}  \\
 		2* & 8 & 13 & 66.27 & 32.27 & Sec.~\ref{sec:global} \\
 		3 & 11 & 36 & 0.922  & 0.408 & Sec.~\ref{sec:nitrogen} \\
 		4** & 31 & 108 & 2.32 & 0.789 & Sec.~\ref{sec:nitrogen} \\
		\hline
	\end{tabular}
	\label{tab:crane_benchmark}
		\begin{tablenotes}\footnotesize
	\item[*] Dynamically runs Bolsig+. 
	\item[**] This example was the same as example 3, but each reaction and species had two duplicates to increase the problem size.
	\end{tablenotes}
	\end{threeparttable}
\end{table}

It is important to note several differences between CRANE and ZDPlasKin which affect computational time. The largest difference is that ZDPlasKin writes Fortran code from the input reactions, so all of the reaction data is compiled prior to runtime. In contrast, CRANE parses all reaction data at runtime, causing a non-negligible portion of computational time to be spent on overhead costs. Additionally, ZDPlasKin is optimized Fortran built specifically for solving 0D reaction networks, while the MOOSE framework is generalized to solve multidimensional PDEs. The benefit of CRANE is that it may be directly coupled into multidimensional models, and as such its raw computational speed is slower than that of ZDPlasKin. 

Another factor that impacts computational speed is dynamically running Bolsig+, which is treated differently by ZDPlasKin and CRANE. This is clearly displayed in the computational time required in the global argon chemistry model (Table~\ref{tab:crane_benchmark}, Test 2). The largest difference is that the Bolsig+ library is compiled into ZDPlasKin but only called as needed by CRANE, which incurs a performance penalty. ZDPlasKin also includes a caching mechanism to improve computational time which does not exist in CRANE, and the inputs to Bolsig+ (number of grid points, convergence tolerance) may be different between the two simulations. All of these factors may cause a significant difference in computational speed between the two models.

\section{Conclusions}
\label{sec:conclusions}

The MOOSE finite element framework is an ideal candidate for plasma simulations due to its inherent scalability and ability to either fully or loosely couple multiple codes together for multiphysics simulations. In this work we have introduced the open source plasma chemistry software, CRANE, in order to further advance the plasma simulation capabilities of the framework. CRANE has shown good agreement with existing ODE solvers and has demonstrated the capability of solving reaction networks, and the framework's ability to run external software was used to externally couple Bolsig+. The ability to compile CRANE into existing MOOSE software was demonstrated with the plasma transport software Zapdos, which was applied to a 1D atmospheric pressure discharge including a network of 13 reactions. 

\section*{Acknowledgements}

This material is based upon work supported by the National Science Foundation under Grant No. 1740310. CRANE is built as part of the MOOSE framework developed at Idaho National Laboratory. The MOOSE team has been an invaluable resource during the development of CRANE. Ternary plots made in Python 2.7 with \texttt{python-ternary} \cite{Harper2015}.





\bibliographystyle{elsarticle-num}
\bibliography{cpc_paper.bib}

\begin{thebibliography}{10}
\expandafter\ifx\csname url\endcsname\relax
  \def\url#1{\texttt{#1}}\fi
\expandafter\ifx\csname urlprefix\endcsname\relax\def\urlprefix{URL }\fi
\expandafter\ifx\csname href\endcsname\relax
  \def\href#1#2{#2} \def\path#1{#1}\fi

\bibitem{Ratovitski2014}
E.~A. Ratovitski, X.~Cheng, D.~Yan, J.~H. Sherman, J.~Canady, B.~Trink,
  M.~Keidar, {Anti-Cancer Therapies of 21st Century: Novel Approach to Treat
  Human Cancers Using Cold Atmospheric Plasma}, Plasma Processes and Polymers
  11 (2014) 1128--1137.
\newblock \href {http://dx.doi.org/10.1002/ppap.201400071}
  {\path{doi:10.1002/ppap.201400071}}.

\bibitem{Ito2018}
M.~Ito, J.-S. Oh, T.~Ohta, M.~Shiratani, M.~Hori, {Current status and future
  prospects of agricultural applications using atmospheric-pressure plasma
  technologies}, Plasma Processes and Polymers 15 (2018) e1700073.
\newblock \href {http://dx.doi.org/10.1002/ppap.201700073}
  {\path{doi:10.1002/ppap.201700073}}.

\bibitem{Zhang2017a}
H.~Zhang, X.~Li, F.~Zhu, K.~Cen, C.~Du, X.~Tu, {Plasma assisted dry reforming
  of methanol for clean syngas production and high-efficiency CO2 conversion},
  Chemical Engineering Journal 310 (2017) 114--119.
\newblock \href {http://dx.doi.org/10.1016/j.cej.2016.10.104}
  {\path{doi:10.1016/j.cej.2016.10.104}}.

\bibitem{Xiong2012}
Z.~Xiong, E.~Robert, V.~Sarron, J.-M. Pouvesle, M.~J. Kushner, {Dynamics of
  ionization wave splitting and merging of atmospheric-pressure plasmas in
  branched dielectric tubes and channels}, J. Phys. D: Appl. Phys 45 (2012)
  275201.
\newblock \href {http://dx.doi.org/10.1088/0022-3727/45/27/275201}
  {\path{doi:10.1088/0022-3727/45/27/275201}}.

\bibitem{Gaston2009}
D.~Gaston, C.~Newman, G.~Hansen, D.~Lebrun-Grandi{\'{e}}, {MOOSE: A parallel
  computational framework for coupled systems of nonlinear equations}, Nuclear
  Engineering and Design 239 (2009) 1768--1778.
\newblock \href {http://dx.doi.org/10.1016/j.nucengdes.2009.05.021}
  {\path{doi:10.1016/j.nucengdes.2009.05.021}}.

\bibitem{Gaston2015}
D.~R. Gaston, C.~J. Permann, J.~W. Peterson, A.~E. Slaughter, D.~Andr{\v{s}},
  Y.~Wang, M.~P. Short, D.~M. Perez, M.~R. Tonks, J.~Ortensi, L.~Zou, R.~C.
  Martineau, {Physics-based multiscale coupling for full core nuclear reactor
  simulation}, Annals of Nuclear Energy 84 (2015) 45--54.
\newblock \href {http://dx.doi.org/10.1016/j.anucene.2014.09.060}
  {\path{doi:10.1016/j.anucene.2014.09.060}}.

\bibitem{Mangeri2015}
J.~Mangeri, O.~Heinonen, D.~Karpeyev, S.~Nakhmanson, {Influence of Elastic and
  Surface Strains on the Optical Properties of Semiconducting Core-Shell
  Nanoparticles}, Physical Review Applied 4 (2015) 014001.
\newblock \href {http://dx.doi.org/10.1103/PhysRevApplied.4.014001}
  {\path{doi:10.1103/PhysRevApplied.4.014001}}.

\bibitem{Peterson2018}
J.~W. Peterson, A.~D. Lindsay, F.~Kong, {Overview of the Incompressible
  Navier–Stokes simulation capabilities in the MOOSE Framework}, Advances in
  Engineering Software 119 (2018) 68--92.
\newblock \href {http://dx.doi.org/10.1016/j.advengsoft.2018.02.004}
  {\path{doi:10.1016/j.advengsoft.2018.02.004}}.

\bibitem{Lindsay2016}
A.~Lindsay, D.~B. Graves, S.~C. Shannon, {Fully coupled simulation of the
  plasma liquid interface and interfacial coefficient effects}, J. Phys. D:
  Appl. Phys 49 (2016) 235204.

\bibitem{Flitti2009}
A.~Flitti, S.~Pancheshnyi, {Gas heating in fast pulsed discharges in N 2-O 2
  mixtures}, Eur. Phys. J. Appl. Phys 45 (2009) 21001.
\newblock \href {http://dx.doi.org/10.1051/epjap/2009011}
  {\path{doi:10.1051/epjap/2009011}}.

\bibitem{Pancheshnyi2008}
S.~Pancheshnyi, B.~Eismann, G.~Hagelaar, L.~Pitchford,
  \href{https://www.zdplaskin.laplace.univ-tlse.fr/}{{Computer code ZDPlasKin}}
  (2008).
\newline\urlprefix\url{https://www.zdplaskin.laplace.univ-tlse.fr/}

\bibitem{Kee1996}
R.~J. Kee, F.~M. Rupley, E.~Meeks, J.~A. Miller, {CHEMKIN-III: A FORTRAN
  CHEMICAL KINETICS PACKAGE FOR THE ANALYSIS OF GAS-PHASE CHEMICAL AND PLASMA
  KINETICS}, Tech. rep., Sandia National Labs., Livermore, CA (1996).

\bibitem{Stark1999}
R.~H. Stark, K.~H. Schoenbach, {Direct current glow discharges in atmospheric
  air ARTICLES YOU MAY BE INTERESTED IN}, Appl. Phys. Lett 74 (1999) 3770.
\newblock \href {http://dx.doi.org/10.1063/1.124174}
  {\path{doi:10.1063/1.124174}}.

\bibitem{Pancheshnyia}
S.~Pancheshnyi,
  \href{https://www.zdplaskin.laplace.univ-tlse.fr/micro-cathode-sustained-discharged-in-ar/}{{Micro-cathode
  sustained discharged in Ar}}.
\newline\urlprefix\url{https://www.zdplaskin.laplace.univ-tlse.fr/micro-cathode-sustained-discharged-in-ar/}

\bibitem{Hagelaar2005}
J.~M. Hagelaar, L.~C. Pitchford, {Solving the Boltzmann equation to obtain
  electron transport coefficients and rate coefficients for fluid models},
  Plasma Sources Sci. Technol. 14~(14) (2005) 722--733.
\newblock \href {http://dx.doi.org/10.1088/0963-0252/14/4/011}
  {\path{doi:10.1088/0963-0252/14/4/011}}.

\bibitem{Rehman2018}
T.~Rehman, \href{www.tue.nl/taverne}{{Studies on plasma-chemical reduction}},
  Phd, Technische Universiteit Eindhoven (2018).
\newline\urlprefix\url{www.tue.nl/taverne}

\bibitem{Pancheshnyi}
S.~Pancheshnyi,
  \href{https://www.zdplaskin.laplace.univ-tlse.fr/external-profiles-of-electron-density-and-electric-field/}{{External
  profiles of electron density and electric field}}.
\newline\urlprefix\url{https://www.zdplaskin.laplace.univ-tlse.fr/external-profiles-of-electron-density-and-electric-field/}

\bibitem{Eliseev2016}
S.~I. Eliseev, A.~A. Kudryavtsev, H.~Liu, Z.~Ning, D.~Yu, A.~S. Chirtsov,
  {Transition From Glow Microdischarge to Arc Discharge With Thermionic Cathode
  in Argon at Atmospheric Pressure}, IEEE Transactions on Plasma Science
  44~(11) (2016) 2536--2544.
\newblock \href {http://dx.doi.org/10.1109/TPS.2016.2557587}
  {\path{doi:10.1109/TPS.2016.2557587}}.

\bibitem{BSRDatabase}
{BSR Database}, \href{www.lxcat.net}{argon{\_}cs{\_}eliseev}.
\newline\urlprefix\url{www.lxcat.net}

\bibitem{Pancheshnyib}
S.~Pancheshnyi,
  \href{https://www.zdplaskin.laplace.univ-tlse.fr/two-reaction-test-case-start-here/}{{Two-reaction
  test case}}.
\newline\urlprefix\url{https://www.zdplaskin.laplace.univ-tlse.fr/two-reaction-test-case-start-here/}

\bibitem{Harper2015}
M.~Harper, B.~Weinstein, C.~Simon, W.~Morgan, V.~Knight, N.~Swanson-Hysell,
  M.~Evans, M.~Greco, G.~Zuidhof, {python-ternary: Ternary Plots in Python.}
  (2015).
\newblock \href {http://dx.doi.org/10.5281/ZENODO.2628066}
  {\path{doi:10.5281/ZENODO.2628066}}.

\end{thebibliography}








\end{document}